\documentstyle[12pt]{article}
\begin{document}
\begin{center}
{\Large \bf
{Preferred Frame in Brane World}
}\\
\vskip 0.5cm
{ Merab GOGBERASHVILI } \\
\vskip 0.3cm
{\it
{Andronikashvili Institute of Physics}\\
{6 Tamarashvili Str., Tbilisi 380077, Georgia}\\
{(E-mail: gogber@hotmail.com)}} \\
\vskip 0.5cm

{\Large \bf Abstract}\\ 
\quotation 
{\small In the framework of brane models the postulates of special relativity 
theory is revised. It is assumed that there exists preferred frame and relativity 
principle is violated on the brane. Because of trapping any moving object on the 
brane is really accelerated and the formulas for gravitational contraction of the 
intervals (containing the escape speed) appears to be equivalent to ordinary 
Lorentz ones.

}
\endquotation
\end{center}

In the present paper we attempt to revise special relativity theory for the brane 
models. Let us start from the simple observation that ordinary special relativity 
globally holds only for the case of single (1+3)-brane and not preserved even for 
the often-considered model of the flat 3-brane with the orthogonal time direction. 
Also introduction of any nontrivial topology of the brane leads to the existence of 
a preferred frame of reference in our world \cite{Pe}. 

We want to show how it is possible to introduce the preferred frame in 
non- contradictive way on the example of the model where our world is considered as 
a single spherical 3-brane expanding in multi-dimensions and matter is confined on 
it by gravitation \cite{G}. In this model two observed facts of modern cosmology, 
the isotropic runaway of galaxies and the existence of a preferred frame of the 
relict background radiation, have the obvious explanation. Note, however, that it 
is not necessary to link directly our considerations to the shell-universe model, 
analogously that for the understanding of the expansion of the universe the extra 
dimensions only help to imagine the picture. 

Let us make several statements:

{\it a). There exists the privileged system of reference and relativity principle
 is violated on the brane. }

There are several reasons to introduce the concept of non-empty vacuum and 
preferred frame in physics. Without such an 'ether' the standard model of 
elementary particles could not be written. Another thing is that for 
consistent formulation of relativistic quantum mechanics may be necessary 
a preferred frame at the fundamental level \cite{BE}, since some correlation 
experiments violate locality \cite{As}. Also in relativistic case the 
satisfactory position operator has not been found up to now.

The negative result of the Michelson-Morley experiment does not rule out the 
preferred frame. Lorentz-type symmetries are just the result of the existence of 
critical velocity and appear even in fluid dynamics. For example, the equation 
for the velocity of sound waves has the form \cite{lamb}:
\begin{equation} \label{sound}
\frac{\partial ^2 u}{\partial t^2}~-~c_s^2~\frac{\partial ^2 u}{\partial x^2}
= 0 ~,
\end{equation}
where $c_s$ is maximal sound speed in the uniform media. Only the case with 
Galilean transformations is excluded, because they do not lead to an invariant 
two-way average velocity of light, which can be measured with only one clock 
\cite{RJ}. The one-way velocity of light cannot be determined experimentally 
\cite{Wi}, since there is a freedom in the definition of the coordinate time and 
one needs two distant clocks, which are synchronized by means of light itself, 
whose velocity is postulated. 

The brane in general breaks down the symmetry of the bulk space-time. 
Consideration of the particular case of the spherical brane means that we are 
introducing privileged frame even for the shell observer (where shell does not 
rotate) and relativity principle is violated \cite{Pe}. 

According to Einstein the fact that Universe stays in a frame rather then in 
another one, does not necessary imply a breakdown of the relativity principle. Of 
course it is possible not to take into account observed cosmological privileged 
frame, however, we know only one Universe and it is a philosophical problem only.
 
{\it b). We use both Mach's and equivalence principles stating that the brane 
determines the inertial and gravitational properties of a body on the brane.}

Einstein's equations, which admit curved vacuum and also asymptotically flat 
solutions, do not incorporate Mach's principle (Einstein rejected it because 
of appearing disagreement with the equivalence principle). We accept the 
formulation of the Mach principle that an isolated body in empty space has no 
inertia, or if you take away all matter, there is no more space \cite{BP}. In 
order to use Mach's principle and still keep the equivalence principle we assume 
that the brane determines both inertial and gravitational properties of a 
4-dimensional body and mass is the measure of interaction of matter with the brane.
 It means, that if we remove the brane both gravitational and inertial masses of 
the body will be zero. Somehow Higgs mechanism must be connected with the trapping 
gravitational potential, since it is known that in Kaluza-Klein models massless 
particles in the bulk space-time are visible as massive on the brane.

{\it c). All clocks with the respect of the preferred frame on the brane can be 
synchronized simultaneously.}

Key point of our approach is to reject relativity principle by introducing 
preferred frame but keep local Poincar\'e invariance for field equations. This is 
possible if we use the coordinate transformations with the non-invariant one-way 
velocity of light, different from the Lorentz ones.

It is known that there are infinite number of possible relativistic 
transformations of coordinates which give the same experimental result, since 
simultaneity and light velocity invariance depend on the way of clocks 
synchronization. For the models with preferred frame instead of the Lorentz 
transformations one can use so-called Ives-Tangherlini formulae \cite{PF} 
\begin{equation} \label{IT} 
t' = t\sqrt{1 - \frac{v^2}{c^2}}~, ~~~x' = \frac{x - vt}{\sqrt{1 - v^2/c^2}}~, 
~~~y' = y~, ~~~ z' = z~, 
\end{equation}
where $(t,x,y,z)$ denotes the coordinates of the privileged frame. The formulae 
(\ref{IT}) imply absolute simultaneity because time transformation does not 
contain $x$-coordinate. However, two events simultaneous for two observers, are 
detected at different time because of the clock slowing down and there is no 
longer light speed invariance 
\begin{equation} \label{c'} 
\frac{1}{c'} = \frac{1}{c} + \frac{v}{c^2} \cos{\theta}~, 
\end{equation} 
where $\theta$ is the angle between $v$ and the light direction in the moving 
system and $c$ is the maximum possible speed for any direction.

It seems that, except of consistent formulation of relativistic quantum theory 
\cite{Re}, so-called clock hypothesis (which states that the rate of an 
accelerated clock is identical to that of the instantaneously co-moving inertial 
frame \cite{Rin}) also requires $x$-independent time transformations (\ref{IT}) 
\cite{Sel}. Clock hypothesis first used by Einstein was confirmed in the muons 
decay experiment in the CERN, where the muons had an acceleration, but their 
time-decay was only due to velocity \cite{Bail}.

We want to choose transformations (\ref{IT}) because in our model all 'clocks' in 
the rest frame of the shell, in which light behaves isotropically, are synchronized 
from the moment of nucleation of the 4-dimensional shell-Universe (what can be 
consider as the big bang for the observer on the brane).

{\it d). In the bulk space-time there is no limit of speed, or it can exceed 
velocity of light on the brane (corresponding to the escape speed of trapped 
matter).}

The assumption that $c$ is not maximal velocity is already used in inflationary 
cosmology considering faster then light expanding Universe. So, in 
multi-dimensional models it is natural to admit superluminary velocities in the 
bulk space-time.

In the case of the spherical shell it is easy to express the maximal possible 
velocity on the brane (the escape speed) by the multi-dimensional Newton's 
gravitational potential of the shell 
\begin{equation} \label{c} 
c = \sqrt{\frac{2m}{M^{d-2}R^{d-3}}}~, 
\end{equation} where $M$ is the fundamental scale, $R$ and $m$ are the radius 
and total mass of the shell respectively and $d$ is the number of space-time 
dimensions. As usual, escape speed for spherical shell is independent of the mass 
of the moving object and of the direction of its velocity.

{\it e). Any moving object on the brane is accelerated and one must observe 
contraction of space-time intervals by gravitational 'ether'.}

What is peculiar of special relativity is the longitudinal contraction of 
material objects and the slowing down of the moving clock rates. From (\ref{IT}) 
follows that slowing of moving clocks is an absolute effect, since, as seen by 
moving observer, clocks in preferred frame run faster. But a meaningful 
comparison of rates implies that a clock at the preferred frame must be compared 
with the clock at two different points of the moving system. The result is dependent 
on the synchronization and therefore on the one-way velocity of light, which for 
the case of transformations (\ref{IT}) is given by the formula (\ref{c'}). So, 
if one can not fix the one-way velocity of light the difference with ordinary 
Lorentz transformations is more apparent then real.

In the case of spherical Universe motion with constant velocity on the shell 
means uniform circular motion for the multi-dimensional inertial observer. Thus, 
uniform velocity on the shell is mathematical abstraction and all moving systems 
on the shell really are accelerated. As a result, a freely moving object in our 
Universe will constantly lose momentum and energy (similar to the model 
\cite{Gho}), in contrast with the conventional physics in which an object moving 
with a constant velocity is not subjected to any force. Since gravity and the 
acceleration motion are locally indistinguishable, fictitious centrifugal force 
for the brane observer can be described like gravity. This means that for any 
moving object on the shell one must introduce speed-depending 'gravitational' 
potential.

As it is known in gravitational field intervals are contracted. Usually it is 
assumed that relativistic slowing of the moving clocks and slowing of the clocks 
in lower gravitational potential are different effects, and general formula for 
the contraction of time interval 
\begin{equation} \label{phi}
dt' = dt\sqrt{1 + \frac{2\phi}{c^2} - \frac{v^2}{c^2}}
\end{equation}
includes both the gravitational potential $\phi$ and the velocity (see, for 
example \cite{Fe}).

We assume that contraction of intervals by the brane gravitational potential is 
absolute effect and both terms in (\ref{phi}) have the same nature. It means 
that a clock is stopped in both cases, when its velocity is equal to $c$ or when its 
total energy is zero. The gravitational potential on an object confined on the 
brane is not zero even in the case we call 4-dimensional vacuum. The total 
energy of a particle on the shell is negative because of trapping potential and 
is equal to zero for the particle with the escape speed $c$.

For the contraction of the time interval we write  
\begin{equation} \label{a}
dt' = dt\sqrt{1 - \frac{a}{A}}~, 
\end{equation}
where $a$ is acceleration of moving object on the shell and $A$ is maximal 
acceleration which corresponds to the value of gravitational trapping energy 
(for another model introducing maximal acceleration see \cite{Ca} and 
references therein). 

In the shell-Universe model the moving observer on the shell must detect 
centripetal acceleration, which for any dimension depends on the square of the 
velocity
\begin{equation} \label{centr}
a = \frac{v^2}{R} ~,
\end{equation}
where $R$ is the radius of the shell. Thus for the uniform motion from (\ref{a}) 
we receive ordinary Lorentz formula for the time interval contraction in the 
moving system
\begin{equation} \label{ds}
dt' = dt\sqrt{1 - \frac{v^2}{c^2}}~, 
\end{equation}
where $c$ is escape velocity from the shell. Note that for the brane with arbitrary 
topology one must introduce an extra assumption about the velocity square depended 
acceleration (\ref{centr}), which naturally arises in the shell-universe model.

So, we claim certain unification of the special and general relativity theories 
based on observation that the Lorentz contraction of intervals is also the 
gravitational effect of the brane and thus come back from Einstein's interpretation 
of free space-time to Poincar\'e's dynamical 'ether'.

{\bf Acknowledgements:} Author would like to acknowledge the hospitality extended 
during his visits at the Abdus Salam International Centre for Theoretical Physics 
where this work was done.


\end{document}